\documentclass[aps,prb,twocolumn,groupedaddress,showpacs]{revtex4}

\usepackage{graphicx}
\usepackage{lmodern} 
\usepackage{multirow}
\usepackage{amsmath}
\usepackage{amsthm}
\usepackage{amsfonts}
\usepackage{color}

\providecommand{\bra}[1]{\left \langle {#1} \right | }
\providecommand{\ket}[1]{\left | {#1} \right \rangle}

\providecommand{\kett}[2]{\left \lvert {#1},{#2} \right \rangle}

\providecommand{\kettt}[3]{\left \lvert {#1},{#2},{#3} \right \rangle}

\providecommand{\ketfive}[5]{\left \lvert {#1},{#2},{#3},{#4},{#5} \right \rangle}

\DeclareSymbolFont{bbold}{U}{bbold}{m}{n}
\DeclareSymbolFontAlphabet{\mathbbold}{bbold}

\begin{document}

\title{In-plane uniaxial magnetic anisotropy in Ga(Mn)As due to local lattice distortions around $\rm{Mn^{2+}}$ ions}

\author{Hemachander Subramanian, J. E. Han}
\affiliation{University at Buffalo, State University of New York, Buffalo, NY, 14226}

\date{\today}

\begin{abstract}
We theoretically investigate the interplay between local lattice distortions around $\rm{Mn^{2+}}$ impurity ion and its magnetization, mediated through spin-orbit coupling of hole. We show that the tetrahedral symmetry around $\rm{Mn^{2+}}$ ion is spontaneously broken in the paramagnetic regime. Modest local lattice distortions around the $\rm{Mn^{2+}}$ ion, along with the growth strain, stabilize magnetization along $\langle 110 \rangle$ directions, in the ferromagnetic regime. We provide a possible explanation for the experimentally observed in-plane uniaxial magnetic anisotropy seen in this system by taking into account the lack of top-down symmetry in this system. 
\end{abstract}


\maketitle

\section{Introduction}
The archetypal dilute magnetic semiconductor (DMS), Ga(Mn)As, has been the object of intense 
investigation over the past decade for its technological relevance and is also of fundamental theoretical
interest ( See [1],[2] and references therein).  The ferromagnetic and semiconducting properties 
impart multifunctionality to this material and it is envisaged to be useful as spintronic
material in future computational architectures where data storage and computation could happen 
at the same place. This charge and spin interaction also provides this material with some 
unusual properties such as temperature and hole density dependent magnetic anisotropy \cite{zemen, zunger, sawicki, timm, ohno}. 
Under compressive strain imparted by the substrate, this material shows both an in-plane uniaxial and cubic magnetic ansiotropy, with the former dominating at high temperatures \cite{sawicki,sawicki1,sawicki2,welp,masmanidis,jarrell}. This in-plane uniaxial ansiotropy has been attributed to surface reconstruction or stacking faults \cite{kopecky}. There is also experimental evidence that this uniaxial anisotropy is a bulk property\cite{sawicki2}. In a recent paper\cite{dietl}, Birowska \textit{et al} suggested that this uniaxial anisotropy is due
to different nearest neighbor Mn pair formation energies along $(110)$ and $(\bar{1}10)$ directions during growth, as observed in their \textit{ab initio} calculations. Although there is substantial experimental evidence, there are few microscopic theories that provide a convincing theoretical explanation for this uniaxial anisotropy. 

In this paper, we investigate the contribution of local lattice distortions around $\rm{Mn^{2+}}$ ion impurity to the magnetic
anisotropy of Ga(Mn)As. Although, from crystal field theory perspective, the ground state of $\rm{Mn^{2+}}$ ion in Ga(Mn)As is a pure spin state because of its half-filled $d$ orbitals, it has been shown\cite{okabayashi,okabayashi2} that the $d$ orbitals of the $\rm{Mn^{2+}}$ ion hybridize with $p$ orbitals of neighboring As ions, and this hybridization should bring the strong spin-orbit (SO) effect into the $\rm{Mn^{2+}}$ ion. This opens up the possibility of local lattice distortions coupling to the spin states of $\rm{Mn^{2+}}$, resulting in local structural and magnetic anisotropy. We expect such coupling with local lattice distortions since it has been observed that the spin states of $\rm{Mn^{2+}}$ in Ga(Mn)P and Ga(Mn)N, materials similar to Ga(Mn)As, couple to local lattice distortion modes \cite{hofmann,stroppa}. Also a photoluminescence experiment\cite{fedorych} on Ga(Mn)As reports 
single-ion uniaxial anisotropy even at very low doping concentrations, implying that the origin of uniaxial anisotropy lies at the single ion level. 

\section{The model}

\begin{figure}[!htb]
\begin{center}
\includegraphics[width=3.6in]{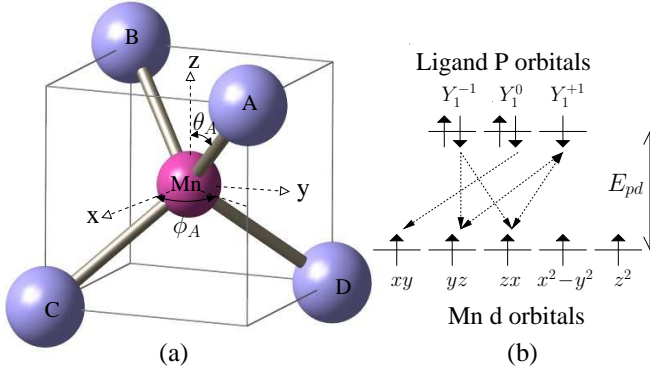}
\end{center}
\caption{ (a) Geometry of the tetrahedral $\rm{MnAs_4}$. The ions denoted as A,B,C and D are As ions. The central $\rm{Mn^{2+}}$ has replaced the
$\rm{Ga^{3+}}$ ion in Ga(Mn)As. (b) Energy level diagram of the $\rm{MnAs}_4$ subsystem. The electronic configurations illustrated 
in the energy level diagram correspond to $\ket{J=\frac{3}{2},J_z=\frac{-3}{2}}$ hole spin state and $\ket{S=\frac{5}{2},S_z=\frac{5}{2}}$ impurity spin state. The energy difference between the impurity $d$ and the acceptor $p$ states, denoted as $E_{pd}$, equals $\epsilon_p-\epsilon_d$. The dotted lines show all the paths with non-zero hopping amplitude (see Table \ref{hopnostrain}) between the impurity ion and the acceptor orbitals for the undistorted $\rm{MnAs}_4$ subsystem. In the undistorted case, hopping connects As $p$ orbitals with the $\rm{Mn^{2+}}$ $d$ orbitals of $t_{2g}$ symmetry.}
\label{unitcell}
\end{figure}

In our model, we consider a single $\rm{Mn^{2+}}$ substitutional impurity ion bonded to the surrounding four 
As ions placed at the vertices of a tetrahedron, as shown in Fig. \ref{unitcell}\cite{ozawa}. We assume that the hole electrostatically bound to the negatively charged (compared to $\rm{Ga}^{3+}$ sublattice) $\rm{Mn^{2+}}$ impurity ion is localized within the $\rm{MnAs}_4$ subsystem, and occupies the As $p$ orbitals. The model explicitly includes only the triply degenerate $4p$ orbitals of the four As ions and the five degenerate $3d$ orbitals of the Mn ion. These two sets of orbitals are hybridized and this $p$-$d$ hybridization is modeled as hopping of electrons between As $p$ orbitals and Mn $d$ orbitals \cite{anderson,larson,bhattacharjee,krstajic,mizokawa,strandberg}. Hopping parameters are worked out using the Slater-Koster approach \cite{slater}. Other orbitals are assumed to be the same as in undoped GaAs. 
The Hamiltonian is 
\begin{equation}\label{fullham}
\begin{aligned}
\mathcal{H} = & \mathcal{H}_a + \mathcal{H}_d + \mathcal{H}_t + \mathcal{E}_{strain}.  \\
\end{aligned}
\end{equation} 
The Hamiltonian consists of four parts. $\mathcal{H}_a$ is for holes in the acceptor states made of $p$ orbitals of As ions, $\mathcal{H}_d$ for electrons localized in the $d$ orbitals of the $\rm{Mn^{2+}}$ impurity ion, $\mathcal{H}_t$ for hopping between the $\rm{Mn^{2+}}$ ion orbitals and the acceptor orbitals, and $\mathcal{E}_{strain}$ for local lattice distortions (treated classically) around the impurity ion. We will discuss each of the four terms individually below. 

This model is motivated by similar descriptions of transition metal ions in II-VI systems by Ley \textit{et al}\cite{ley}, Mizokawa and Fujimori\cite{mizokawa1,mizokawa}, and of Mn ion in III-V systems by Okabayashi \textit{et al}\cite{okabayashi,okabayashi2} and, Ivanov and Krstajic \textit{et al}\cite{ivanov,krstajic}. With the exception of Ivanov and Krstajic, all others treat the DMS systems as five-ion clusters and calculate exchange constants using configuration interaction scheme. These models also assume that the $p$-$d$ interaction between Mn and the holes in the host lattice is independent of hole wave-vector. Modifications and improvements in the current model over the above models are the inclusion of spin-orbit coupling in the acceptor part of the Hamiltonian $\mathcal{H}_a$, introduction of a local strain energy term to take into account local lattice distortions around the $\rm{Mn^{2+}}$ ion, and the use of an intra-atomic Coulomb interaction Hamiltonian $\mathcal{H}_d$\cite{oles} which is rotationally invariant in the spin space.

\subsubsection{Acceptor Hamiltonian $\mathcal{H}_a$} 
The first term in the equation \eqref{fullham} describes the acceptor states in As ions created due to the substitution of a $\rm{Mn^{2+}}$ impurity ion at the position of  $\rm{Ga^{3+}}$ ion. The hole captured in the hydrogenic acceptor state\cite{linnarsson, okabayashi} is assumed to reside in orbitals formed by linear combination of the atomic $p$ orbitals of the four As ions. At low doping concentrations and temperature, it can be reasonably assumed that the hole is at ${\bf k}=0$ $\Gamma$-point\cite{ley,mizokawa1,mizokawa,okabayashi,okabayashi2,ivanov,krstajic}, which implies that the ground state of the hole wavefunction has the same phase at all the four As ions. Thus, of all possible linear combinations of the twelve atomic As-$p$ orbitals (ignoring spin), only three combinations have the same phase and will be used in the modeling below. If we let $\ket{{p_i^n}}$ denote the $p_i$ orbital of $n^{th}$ As ion, with $n$ being one of A, B, C or D, then the three orbitals are
\begin{equation}\label{holeorbitals}
\ket{P_i} = \frac{1}{2}\left ( \ket{p_i^A} + \ket{p_i^B} + \ket{p_i^C}+ \ket{p_i^D} \right ), \hspace{10pt} i=x,y,z. 
\end{equation}
These three (six, if spin is taken into account) degenerate $p$ orbitals are assumed to hold one hole in the ground state of the $\mathcal{H}_a$ Hamiltonian. We take the three $P$ orbitals to be the eigenstates of $L_z$ operator, which are just the $Y_1^{-1}, Y_1^{0}, Y_1^{+1}$ spherical harmonic functions. 
\begin{equation}
\begin{aligned}
\mathcal{H}_a = & \epsilon_p \sum_{i,\sigma} c_{i\sigma}^\dagger c_{i\sigma} - \lambda \mathbf{L}\cdot\mathbf{S}\\ 
 = & \epsilon_p \sum_{i,\sigma} c_{i\sigma}^\dagger c_{i\sigma} - \lambda \left ( \sum_{j} \mathbf{L}_j \right ).\left ( \sum_{k} \mathbf{S}_k \right ). \\ 
\end{aligned}
\end{equation}
In the above equation, $c_{i\sigma}^\dagger$ and $c_{i\sigma}$ are creation and annihilation operators for electrons. The second term
introduces spin-orbit coupling to the electrons, where $\lambda$ denotes the spin-orbit coupling constant. The subscripts $j$ and $k$ in the second term run through the number of electrons, and not orbitals. The many-body basis states and the eigenstates of $\mathcal{H}_a$ for the case of five electrons in the three $P$ orbitals (corresponding to the presence of a single hole) are listed in Appendix A. With the introduction of spin-orbit coupling, the six degenerate many-body states split into four degenerate $J=\frac{3}{2}$ states of energy $5\epsilon_p - \lambda/2 $ and doubly degenerate $J=\frac{1}{2}$ states of energy $5\epsilon_p + \lambda$. Thus in our prameterization, the usual spin-orbit splitting at the $\Gamma$-point $\Delta_{SO} = 3 \lambda/2$. The value of $\lambda$, derived from $\Delta_{SO}=0.341$ eV \cite{cardona}, equals $0.23$ eV. We similarly work out the excited many-body eigenstates with electron numbers other than five. 

\subsubsection{$\rm{Mn^{2+}}$ impurity Hamiltonian $\mathcal{H}_d$} 

The impurity ion Hamiltonian describes the intra-atomic Coulomb interaction within $\rm{Mn^{2+}}$ $d$ orbitals.  
\begin{equation}
\begin{aligned} 
\mathcal{H}_d = & \epsilon_d \sum_{i,\sigma} n_{i\sigma}  + \frac{1}{2}\sum_{\stackrel{i \neq j}{\sigma}} U n_{i,\sigma}n_{j,-\sigma} \\
+ & \sum_i  \left( U+2J \right) n_{i\uparrow}n_{i\downarrow} 
+   \frac{1}{2}\sum_{\stackrel{i \neq j}{\sigma}} \left( U-J \right)  n_{i\sigma}n_{j\sigma} \\
+  & \frac{1}{2}\sum_{\stackrel{i \neq j}{\sigma}} 
J \left[  d^{\dagger}_{i,\sigma} d^{\dagger}_{j,-\sigma} d_{i,-\sigma}  d_{j,\sigma} 
+  d^{\dagger}_{i,\sigma} d^{\dagger}_{i,-\sigma} d_{j,-\sigma} d_{j,\sigma} \right]. \\
\end{aligned}
\end{equation}
In the above equation, $d_{i\sigma}^\dagger$, $d_{i\sigma}$ and $n_{i\sigma}$ denote creation, annihilation and number operators respectively of electron with spin $\sigma$ in the $i^{\rm{th}}$ $d$ orbital of the impurity ion. $U$ and $J$ are the intra-ionic direct and exchange Coulomb interaction parameters, respectively. The widely used Hamiltonian $\mathcal{H}_d$ is rotationally invariant in the spin space of electrons\cite{oles}. We choose the ground state occupancy of the impurity ion to be five, in accordance with the results of numerous experiments \cite{twardowski,linnarsson,okabayashi}. The ground eigenstate degeneracy of the impurity is six, with the eigenstates being the six $S_z$ states of the total spin $S=\frac{5}{2}$. These many-body states are explicitly written down in Appendix B. This degeneracy is the result of the spin rotational symmetry of $\mathcal{H}_d$. This result is also in accordance with Hund's first rule of maximizing the total spin $S$ of the ion.

The ground state energy of the $\rm{Mn}^{2+}$ with five electrons in the $d$ orbitals, denoted by $E(d^5)$, is $5\epsilon_d+10U-10J$. 
The center of gravity of the $\rm{Mn}^{+}$ ion multiplet, with six electrons in the $d$ orbitals, denoted by $E(d^6)$, is $6\epsilon_d+15U-5J$, and
similarly, $E(d^4) = 4\epsilon_d+6U-3J$. In their photo-excitation experiment\cite{okabayashi}, Okabayashi \textit{et al} measured $E(d^6)+E(d^4)-2E(d^5)$ (which, in our parametrization, is $U+12J$) to be $3.5\: \rm{eV}$ (see table 1 of the above reference). Using the above value, and assuming $J=0.2U $\cite{friedel,oles}, we determine the values of $U$ and $J$ to be $1.1 \:\rm{eV}$ and $0.22 \:\rm{eV}$, respectively. Since
the ground state electronic configuration of Mn ion is found\cite{twardowski,linnarsson,okabayashi} to be the stable half-filled $d^5$ configuration, $E(d^5)$ must be smaller than both $E(d^4)$ and $E(d^6)$. Thus, with the above values of $U$ and $J$, the value of $\epsilon_d$ measured from the top of the valence band must be between $-2.7 \: \rm{eV}$ and $-6.2 \: \rm{eV}$. We choose $\epsilon_d-\epsilon_p$ to be $-4.5$ eV, the mean value of the above limits. Many-body eigenstates of $\mathcal{H}_d$ with number of electrons other than five are also worked out similarly. Table \ref{parameters} lists all the parameters used in the model, for ease of reference.

\begin{table}
	\centering
		\begin{tabular}{|c|c|c|c|}
		 \hline
		         
        Parameter & Value & Parameter & Value \\
                 
     \hline\hline    
     
       $\epsilon_d-\epsilon_p$ & $-4.5$ eV & $\lambda$ & $0.23$ eV \cite{cardona} \\
       
     \hline  
       
       $U$ & $1.1$ eV \cite{okabayashi} & $J$ & $0.22$ eV \cite{okabayashi} \\
     
     \hline     
     
       $(pd\sigma)_0$  & 1 eV \cite{okabayashi2} & $(pd\pi)_0$ & -0.46 eV \cite{okabayashi2} \\
     
     \hline
     
      $\alpha$ & 41.19 $\rm{N/m}$ \cite{martins} & $\beta$ & 8.94 $\rm{N/m}$ \cite{martins} \\   
     
     \hline        
     
     \end{tabular}
	\caption{List of parameters used in the model, provided for ease of reference. The table lists, by row, energy of $d$ orbitals with respect to top of the valence band and spin-orbit coupling constant, Coulomb direct and exchange integral	values, Slater-Koster overlap integrals, and, radial and angular strain constants used in VFF model, respectively.} 
	\label{parameters}
\end{table}

\subsubsection{Hamiltonian for local lattice distortion, $\mathcal{E}_{strain}$} 

It is theoretically possible for the $\rm{Mn^{2+}}$ impurity to induce local lattice distortions\cite{demchenko,katsumoto}. 
Because of the spin-orbit coupling of holes and $p$-$d$ hybridization between holes and the Mn spin, the spin states of 
the $\rm{Mn^{2+}}$ impurity ion can couple with local lattice distortions \cite{hofmann,guo,stroppa}. In this model, we treat the 
local lattice distortions classically. We only take into account the distortions of the nearest neighbors of $\rm{Mn^{2+}}$ ion, i.e., the four As ions A,B,C and D (see Fig. \ref{unitcell}). For the five-ion $\rm{MnAs_4}$ sub-system of Ga(Mn)As, $15$ variables are needed to completely specify the distorted configuration. Rotational and translational invariance remove $6$ variables leaving $9$ to completely specify local lattice distortions. These $9$ variables are chosen as follows. Four variables $d_A$, $d_B$, $d_C$ and $d_D$ are used to denote the four Mn-As bond lengths, leaving five variables to specify angular distortions. We choose angles of the bond connecting the Mn and As ion B to be fixed, and allow As ion A to move only in the A-Mn-B plane; i.e., along the $\theta_A$ direction (see Fig. \ref{unitcell}(a)). The other two ions are allowed to freely rotate about the Mn ion in all directions. We have verified that choosing a pair of As ions other than A and B for applying constraints give the same results (with some choices requiring a constraint equation to be solved).

We parametrize the strain energy of the $\rm{MnAs_4}$ sub-system using Keating's valence force field (VFF) model\cite{keating, martins}:

\begin{equation}
\begin{aligned}
\mathcal{E}_{strain} = & 2\sum_{i=A,B,C,D} \frac{3 \alpha d_0^2}{8} \left ( \left(\frac{d_i}{d_0}\right)^2 - 1 \right )^2 \\
       + & \sum_{\stackrel{i,j=A,B,C,D}{i>j}} \frac{3 \beta d_0^2}{8} \left (\frac{\mathbf{d}_i\cdot\mathbf{d}_j}{d_0^2} + \frac{1}{3} \right )^2. \\       
\end{aligned}\label{vff}
\end{equation}
Here, $\textbf{d}_i$ denotes the position vector of the $i^{\rm{th}}$ As ion, and ${d_0}$, the equilibrium Mn-As bond length. The sums run over all the four As ions. The parameters $\alpha$ and $\beta$ used in VFF model are known for GaAs\cite{martins}, and are $41.19 \: \rm{N/m}$ and $8.94 \: \rm{N/m}$, respectively. These values are used to evaluate the prefactors in the above equation. We take $d_0$ to be the equilibrium Ga-As bond length. The prefactor $2$ in front of the radial strain energy term is to approximately include the radial strain energy of the bonds connecting the As ions A,B,C and D to their nearest neighbor Ga ions (next nearest neighbors to $\rm{Mn}^{2+}$) which are to be fixed at their equilibrium positions. This approximation is equivalent to replacing all the Ga-As bonds from a given As ion with a single bond, collinear with, and having the same spring constant as that of the Mn-As bond. The assumption of a fixed Ga sublattice around Mn ion is a good approximation, since, an X-ray absorption spectroscopy study\cite{acapito} of the local structure around $\rm{Mn}^{2+}$ impurity ion in Ga(Mn)As finds that the structural perturbation induced by the impurity ion is very localized. With these  parameters, the prefactors in Eq. \eqref{vff} are $\frac{3 \alpha d_0^2}{8} = 5.7 \: \rm{eV}$ and $\frac{3 \beta d_0^2}{8} = 1.3 \: \rm{eV}$.

It has to be noted that the distortions couple to the electronic degrees of freedom only through the hopping amplitudes between $d$ and $p$ orbitals (see Eqs. \eqref{hopdef2} and \eqref{hopdef3} below). In other words, it is assumed that the acceptor Hamiltonian $\mathcal{H}_a$ is affected by local lattice distortions only through its coupling with the $\rm{Mn^{2+}}$ impurity ion. This assumption can be rationalized by invoking the difference between the strengths of the Mn-As and the Ga-As bonds, derived from elastic constants of \textit{zinc blende} MnAs and GaAs. The spring constant $\alpha$ of the Mn-As bond, derived from the bulk elastic constants of \textit{zinc blende} MnAs\cite{keating}, is about $60\:\%$ of the value of spring constant of Ga-As bond\cite{qi}, implying that the Mn-As bonds are weaker than the Ga-As bonds. Thus local lattice distortions around the $\rm{Mn}^{2+}$ impurity ion can also be approximately modeled as the impurity ion itself moving with respect to four As ions fixed at their equilibrium positions. We verified this by fixing the As ion positions and allowing only the $\rm{Mn}^{2+}$ ion to move within this fixed four As ion ``cage'' and reproduced essentially the same physics. Thus the assumption that the acceptor states are not affected directly by local lattice distortions is physically reasonable. Also, with this assumption, we achieve considerable computational simplification, by removing the need for self-consistent calculation of hole eigenstates in the presence of distortions. This simpler strain model, where the $\rm{Mn}^{2+}$ ion moves with respect to a fixed As ion cage, 
provides a simpler way to visualize the local lattice distortion modes, as will be shown below.

\subsubsection{The hopping Hamiltonian, $\mathcal{H}_t$} 

Finally, the $p$-$d$ hybridization between As $p$ orbitals and $\rm{Mn^{2+}}$ $d$ orbitals is modeled as hopping of electrons 
between the $p$ and $d$ orbitals.

\begin{equation}\label{hopeq}
\begin{aligned}
\mathcal{H}_t = & \sum_{i,j} t_{ij} c_{i\sigma}^\dagger d_{j\sigma} + \text{h.c}.  \\
\end{aligned}
\end{equation} 
In Eq. \eqref{hopeq}, $i$ runs through all the three $P$ orbitals of As ion defined in Eq. \eqref{holeorbitals}, and $j$ runs through the five $d$ orbitals of $\rm{Mn^{2+}}$ ion. The hopping amplitudes $t_{ij}$'s are functions of the local lattice distortion variables and are derived using Slater-Koster formulae\cite{slater}. To illustrate the derivation of these hopping amplitudes, we explicitly evaluate $t_{+1,yz}$, the hopping amplitude between the $Y_1^1$ orbital and $yz$ orbital of $\rm{Mn^{2+}}$, with the latter assumed to be located at the origin of the coordinate system. 

\begin{equation}\label{hopdef1}
\begin{aligned}
& t_{+1 ,yz} \equiv \bra{P_{+1}} V_{latt} \ket{d_{yz}} \\
  = & \frac{-1}{\sqrt{2}} \bra{P_{x}} V_{latt} \ket{d_{yz}} + 
                                                             \frac{i}{\sqrt{2}} \bra{P_{y}} V_{latt} \ket{d_{yz}} \\
          = & \frac{-1}{2\sqrt{2}}\left ( \bra{p_x^A} + \bra{p_x^B} + \bra{p_x^C} + \bra{p_x^D} \right ) V_{latt} \ket{d_{yz}} + \\
            & \frac{i}{2\sqrt{2}}\left ( \bra{p_y^A} + \bra{p_y^B} + \bra{p_y^C} + \bra{p_y^D} \right ) V_{latt} \ket{d_{yz}}. \\                                                  
\end{aligned}
\end{equation} 
If we let $(l_k,m_k,n_k)$ and $d_k$ be the direction cosines and the length of the bond connecting $\rm{Mn^{2+}}$ ion and $k^{\rm{th}}$ As ion,  
with $k=\rm{\{A,B,C,D\}}$, we can write the above equation for the hopping amplitude $t_{+1,yz}$ using Slater-Koster tables as

\begin{equation}\label{hopdef2}
\begin{aligned}
t_{+1,yz} = & \frac{-1}{2\sqrt{2}} \sum^D_{k=A} [ \sqrt{3} l_k m_k n_k \left ( pd\sigma \right )  - 2 l_k m_k n_k \left ( pd\pi \right ) \\
& -\sqrt{3} i m_k^2 n_k \left ( pd\sigma \right )  - i \left ( 1-2 m_k^2 \right ) n_k \left ( pd\pi \right ) ]. \\
\end{aligned}
\end{equation} 
The hopping amplitude is assumed to depend on bond length distortion through modification of hopping parameters $(pd\sigma)$ and $(pd\pi)$ according
to Harrison's rule\cite{harrison}:

\begin{equation}\label{hopdef3}
\begin{aligned}
\left ( pd\sigma \right ) = & \left ( \frac{d_0}{d} \right )^2 \left ( pd\sigma \right )_0  \hspace{10pt}  \text{and} \hspace{10pt}
\left ( pd\pi \right ) =  \left ( \frac{d_0}{d} \right )^2 \left ( pd\pi \right )_0.  \\
\end{aligned}
\end{equation} 
Here, $(pd\sigma)_0$ and $(pd\pi)_0$ are hopping parameters for undistorted Mn-As bonds. We assume $(pd\sigma)_0 = 1 \: \rm{eV}$ \cite{okabayashi2},
and use the relation $(pd\sigma)_0 \approx -2.17 (pd\pi)_0$ \cite{harrison2} to derive $(pd\pi)_0$.  When the $\rm{MnAs_4}$ subsystem is undistorted, the expression for $t_{+1,yz}$ simplfies to
 
\begin{equation}\label{hopdef4}
\begin{aligned}
t_{+1,yz} = & -\sqrt{2} \left [ \frac{(pd\sigma)_0}{3} - \frac{2(pd\pi)_0}{3\sqrt{3}} \right ] \equiv - \frac{t_0}{\sqrt{2}}.\\
\end{aligned}
\end{equation} 
Similarly

\begin{equation}\label{hopdef5}
\begin{aligned}
t_{-1,yz} = & \frac{t_0}{\sqrt{2}} \hspace{10pt}  \text{and} \hspace{10pt}  t_{+1,yz} = & -(t_{-1,yz})^*. \\
\end{aligned}
\end{equation} 
All the fifteen hopping amplitudes between real orbitals for the undistorted $\rm{MnAs_4}$ subsystem are given in table \ref{hopnostrain},
in terms of $t_0$, defined in \eqref{hopdef4}.

\begin{table}
	\centering
		\begin{tabular}{|c||c|c|c|c|c|}
		 \hline
		         
        $V_{latt}$ & $\ket{d_{xy}}$ & $\ket{d_{yz}}$ & $\ket{d_{zx}}$ & $\ket{d_{x^2-y^2}}$ & $\ket{d_{3z^2-r^2}}$ \\
                 
     \hline\hline
     
     $\bra{P_x}$ & $0$ & $-t_0$ & $0$ & $0$ & $0$ \\
     
     \hline
      
     $\bra{P_y}$ & $0$ & $0$ & $-t_0$ & $0$ & $0$ \\
     
     \hline
     
     $\bra{P_z}$ & $-t_0$ & $0$ & $0$ & $0$ & $0$ \\
     
     \hline

     \end{tabular}
	\caption{Hopping amplitudes in the absence of local lattice distortions. First three $d$ orbitals of $t_{2g}$ symmetry
	couple with atleast one $P$ orbital in the presence of lattice potential $V_{latt}$, whereas the last two $d$ orbitals of $e_g$ symmetry
	do not couple with any of the $P$ orbitals.}
	\label{hopnostrain}
\end{table}

\section{Calculational details and analysis}

We solve the Hamiltonian $\mathcal{H}$ using second-order degenerate perturbation theory, by expanding the ground state energy upto second-order in the hopping amplitudes $t_{i,j}$'s\cite{mizokawa,okabayashi}. The degenerate ground state manifold of $\mathcal{H}_0=\mathcal{H}_a+\mathcal{H}_d$ is defined as follows. The ground eigenstates of the acceptor Hamiltonian $\mathcal{H}_a$ are the four degenerate $\ket{J=\frac{3}{2}, J_z=\pm\frac{3}{2},\pm\frac{1}{2}}$ total angular momentum eigenstates given in Appendix A. We ignore the $J=\frac{1}{2}$ split-off band eigenstates. The ground eigenstates of the impurity Hamiltonian $\mathcal{H}_d$ are the six degenerate $\ket{S=\frac{5}{2},S_z=\pm\frac{5}{2},\pm\frac{3}{2},\pm\frac{1}{2}}$ eigenstates, given in Appendix B. Thus the Hilbert space in which we will work is the 24-dimensional product space of the two sets of eigenstates mentioned above, i.e., $\ket{J=\frac{3}{2}, J_z=\pm\frac{3}{2},\pm\frac{1}{2}} \otimes \ket{S=\frac{5}{2},S_z=\pm\frac{5}{2},\pm\frac{3}{2},\pm\frac{1}{2}}$. Let $E_0$ be the energy of these 24 degenerate states.

All the terms arising from first-order perturbation expansion around $\mathcal{H}_t$ vanish trivially, since $\mathcal{H}_t$ changes the total number of electrons inside the $\rm{Mn}^{2+}$ ion. We treat paramagnetic case by representing the second-order perturbation operator $\mathcal{H}_t \frac{1}{(E_0-\mathcal{H}_0)} \mathcal{H}_t$ within the 24-dimensional Hilbert space spanned by the basis vectors mentioned above. We diagonalize the $24 \times 24$ Hamiltonian matrix to find the ground state eigenvalue, add strain energy, and minimize the total ground state energy with respect to the distortion variables to find the minimum energy configuration of the $\rm{MnAs_4}$ subsystem. We have verified that the Hamiltonian matrix and its eigenvectors preserve time-reversal and total spin rotational symmetry, in the absence of any magnetic field or local lattice distortions. 

When the DMS system is ferromagnetic, the $\rm{Mn}^{2+}$ ion is magnetized along a particular direction, and we replace its spin operator by a fixed vector. We treat the ferromagnetic case by projecting the 24-dimensional matrix down to a 4-dimensional matrix with only the hole spin degrees of freedom included. We then minimize the ground state energy ( with strain energy included) with respect to distortion variables. Finally, growth-induced strain effects are taken into account similarly, by reducing the Hilbert space size and working within either light-hole or heavy-hole sub-bands, depending upon the type of strain. 

\subsubsection*{Paramagnetic case:} 

In the absence of hopping and distortions, the paramagnetic ground state degeneracy of the many-body states is $24$ ($6$ from $S=5/2$ states of Mn and $4$ from $J=3/2$ states of hole). With hopping, this $24$-fold degenerate ground state split into $3,5,7$ and $9$-fold degenerate states. This splitting can be understood using the total angular momentum of the $\rm{Mn^{2+}}$-hole complex. The quantum number corresponding to the total angular momentum  $\mathbf{F} = \mathbf{J} + \mathbf{S}$ can take the values $1,2,3$ and $4$ with the degeneracies of $3,5,7$ and $9$ respectively, and the ground state manifold is $F=1$, because of the anti-ferromagnetic coupling\cite{ohno1} between holes and $\rm{Mn^{2+}}$. Thus we observe that the total spin rotational symmetry of the entire $\rm{MnAs}_4$ subsystem is not broken with the introduction of hopping and the calculated eigenstates are also the eigenstates of the total angular momentum operator $\mathbf{F}$. A heuristic reason for this total spin rotational invariance, in the special case of $F=1$, is provided in Appendix C. This $24$-dimensional Hamiltonian matrix, whose eigenvectors are also eigenvectors of the total angular momentum operator $\mathbf{F}$ and its z-component $F_z$, is then identical, apart from a constant spin-independent diagonal term, to the following Hamiltonian involving only the spin variables: 
\begin{equation}\label{spinint}
\begin{aligned}
\mathcal{H}_{\rm{eff}} = J_0 \hat{\textbf{J}}\cdot\hat{\textbf{S}}. 
\end{aligned}
\end{equation} 
In the above equation, the components of $\hat{\textbf{J}}$ ($\hat{\textbf{S}}$) refer to angular momentum matrices for spins $3/2$ ($5/2$), with $3/2$ ($5/2$) factored out from the matrices, and $J_0$ is the exchange constant.   The interaction between $\rm{Mn^{2+}}$ spin and the hole spin is anti-ferromagnetic, and is of Heisenberg type. With our choice of the parameters listed in table \ref{parameters}, $J_0 = 0.76\: \rm{eV}$\cite{mynote2}, close to the accepted value\cite{okabayashi} of $1.2 \pm 0.2 \: \rm{eV}$. This effective spin Hamiltonian is a direct consequence of the conservation of the total angular momentum $\mathbf{F}$ and the second-order perturbation approximation. 

In the paramagnetic case, the time-reversal symmetry of the Hamiltonian is not broken, and we expect the spin expectation values of the hole and the $\rm{Mn^{2+}}$ ion to vanish in the ground state, irrespective of distortions. The ground eigenstate energy minimized with respect to the local lattice distortions is singly degenerate, and is time-reversal symmetric with itself. If we denote the ground state eigenvector as $\ket{\Psi_{GS}}$, then, time-reversal symmetry implies
\begin{equation}\label{timereversal}
\begin{aligned}
\Theta \ket{\Psi_{GS}} = \ket{\Psi_{GS}}.
\end{aligned}
\end{equation}
Here $\Theta$ is the time-reversal operator. If $\textbf{S}$ denote the spin angular momentum operator of either the hole or $\rm{Mn^{2+}}$ ion, then,

\begin{equation}\label{timereversal1}
\begin{aligned}
\Theta \textbf{S} \Theta^{-1} = -\textbf{S}. 
\end{aligned}
\end{equation}
Therefore,

\begin{equation}\label{timereversal2}
\begin{aligned}
 \bra{\Psi_{GS}}\textbf{S} \ket{\Psi_{GS}} = & 0. \\
\end{aligned}
\end{equation}
Thus, even when local lattice distortions are introduced, the spin expectation values of the hole and the $\rm{Mn^{2+}}$ ion is
zero, for the ground eigenstate\cite{mynote}. Distortions break orbital, and through spin-orbit coupling, spin rotational symmetry, but not time-reversal symmetry. Breaking of time-reversal symmetry is needed for the appearance of a magnetic dipole moment. 
The breaking of spin rotational symmetry, due to distortions, is seen at the quadrupole level, with $\langle S_x^2 \rangle \ne \langle S_y^2 \rangle \ne \langle S_z^2 \rangle$, in general, for the ground eigenstate.

\vphantom{10pt}
\paragraph*{Energy minimization with respect to distortion variables:} 

\begin{figure}[!htb]
\begin{center}
\includegraphics[width=4in]{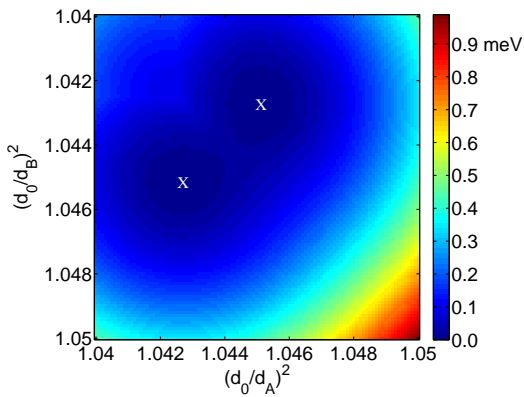}
\end{center}
\caption{An example of symmetry breaking in $\rm{MnAs}_4$ tetrahedron when the ground state energy is minimized with only two radial strain variables, keeping the other two equal and constant and all angular variables fixed at their equilibrium values. Energy is minimum when $d_A$ and $d_B$, the length of bonds connecting Mn with As ions A and B, are different. Two degenerate energy minima, denoted by ``X'' symbols, seen in the plot reflect the permutational symmetry that exists between $d_A$ and $d_B$.}
\label{twobonds}
\end{figure}

When we minimize the ground state energy of $\mathcal{H}$ with respect to all the nine distortion variables, we observe \textit{spontaneous breaking of tetrahedral symmetry} of the As ions around $\rm{Mn^{2+}}$ ion. Minimization of energy with respect to just radial distortion variables, keeping angles of the Mn-As bonds at equilibrium values, result in a configuration where one of the bonds grow shorter ($5.2\%$ less than the equilibrium value $d_0$) than the other three bonds ($4.8\%$ less than $d_0$). The shorter bond can be any of the four bonds, and the system has permutational symmetry with respect to the radial distortion variables. The energy landscape has degenerate energy minima corresponding to all the four permutations of distortions. Figure \ref{twobonds} shows the symmetry breaking when the energy is minimized with only two radial strain variables, say $d_A$ and $d_B$, keeping the other two radial strain variables equal and constant, and keeping all angles at their equilibrium values. The plot shows that the energy is minimal when $d_A$ and $d_B$ are unequal. 

\begin{figure}[!htb]
\begin{center}
\includegraphics[width=2.8in]{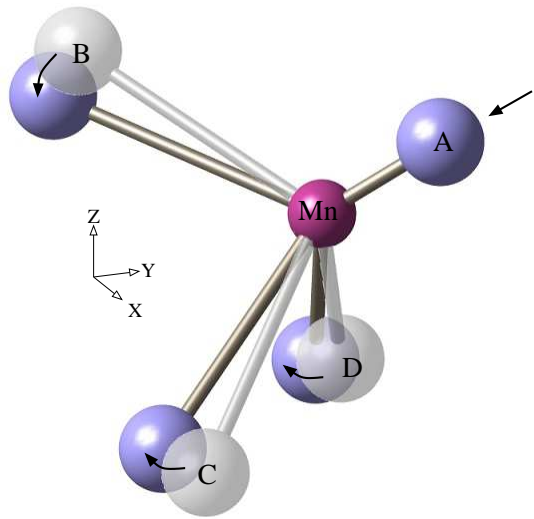}
\end{center}
\caption{Local lattice distortion when the system is in the paramagnetic regime and at one of the four degenerate energy minima, exaggerated for clarity. One of the bonds, Mn-A, becomes shorter by $0.4 \%$ than the other three bonds, while the other three bonds move away from the shorter bond by $0.4$ degrees. The tetrahedral symmetry around the $\rm{Mn}^{2+}$ ion is spontaneously broken. In the alternate simplified strain model in which the only strain variable is the position vector of the $\rm{Mn}^{2+}$ impurity ion, the $\rm{Mn}^{2+}$ ion moves towards the fixed As ion A. Configurations at other degenerate energy minima have the same structure but with the As ion labels permuted.}
\label{shortdistortion}
\end{figure}

Minimization of the ground state energy with only angular distortion variables result in a broken symmetry configuration where three bonds move away from the fourth bond (by about $0.4$ degrees) while keeping their relative angles the same. Again, permutational symmetry holds and the singled-out bond can be any one of the four bonds, with the energy surface containing four degenerate energy minima corresponding to the four possible angular distortion modes. When both radial and angular distortion variables are included, the three longer bonds move away from the shorter one. Figure \ref{shortdistortion} shows the resultant configuration of the $\rm{MnAs}_4$ subsystem. This mode of local lattice distortion is equivalent to 
$\rm{Mn}^{2+}$ ion moving towards the fixed As ion A in the simplified strain model introduced above, in which the As ions are fixed at their equilibrium positions and the $\rm{Mn}^{2+}$ ion is allowed to move within the As ion cage. This simplified picture of $\rm{Mn}^{2+}$ ion displacement is approximately equivalent to the full five-ion distortion picture.
With distortion, the triply denegerate $F=1$ ground state manifold splits into a self-time-reversal symmetric ground state singlet and an excited state Kramers doublet. The ground state ceases to be an exact eigenstate of the total angular momentum operator $\textbf{F}$, and gains a small amount of higher total angular momentum, apart from $F=1$.  

So far, we have not included the growth strain along the $\left (001 \right )$ direction in the model. From experiments, we know that in-plane magnetization is the result of compressive strain exerted by substrates such as GaAs on the Ga(Mn)As film\cite{shen,liu}. This strain splits
the $J=\frac{3}{2}$ manifold into light-hole ($J_z=\pm\frac{1}{2}$) and heavy-hole ($J_z=\pm\frac{3}{2}$) Kramers doublets, with the light-hole doublet remaining lower in energy\cite{jungwirth,zemen,sawicki,sawicki1,sawicki2}. To include the effect of compressive growth strain, we assume that the holes populate only the light-hole band, and hence include just the light-hole $J_z=\pm\frac{1}{2}$ doublet in the model, and neglect the heavy-hole doublet altogether. This results in a reduction of the size of our Hilbert space from $24$ to $12$. With the inclusion of growth strain, minimization of ground state energy with respect to local lattice distortion result in equal shortening of the four bonds with no angular distortions. Thus, symmetry-breaking vanishes when only light-holes are considered, in the case of unpolarized $\rm{Mn}^{2+}$ ion spin. Realistically, above $T_c$, in the bulk DMS sample, the population of holes occupying the heavy-hole band will not be negligible, and experiments should be able to observe weak local lattice distortions. 
With magnetization of the $\rm{Mn}^{2+}$ ion, distortions reappear, and are coupled to the direction of magnetization, as we will see below.

\subsubsection*{Ferromagnetic case:}  

At very low temperatures $(T \ll T_c)$, the magnetic moments of all the $\rm{Mn}^{2+}$ impurity ions align in a particular direction, rendering the DMS sample magnetized. We model the ferromagnetic case by magnetizing the impurity ion along a direction parametrized by the polar and azimuthal angles $(\Theta,\Phi)$, using rotation operators:

\begin{equation}\label{polarize}
\begin{aligned}
\ket{\chi\left ( \Theta,\Phi \right )} = e^{-iS_z\Phi} e^{-iS_y\Theta} \: [1,0,0,0,0,0]^T.
\end{aligned}
\end{equation}  
In Eq. \eqref{polarize}, $S_y$ and $S_z$ represent angular momentum matrices for spin $\frac{5}{2}$ corresponding to rotations about y and z-axes respectively, and $\ket{\chi\left ( \Theta,\Phi \right )}$ represent the impurity ion spinor polarized along $(\Theta,\Phi)$ direction. 
After the removal of spin degrees of freedom of the $\rm{Mn}^{2+}$ impurity ion by magnetizing it, we are left with only the hole degrees of freedom and the size of Hilbert space reduces further to just four, ignoring the effects of growth strain for now. Thus, we diagonalize the $4 \times 4$ Hamiltonian matrix, whose elements are functions of distortion variables entering through hopping amplitudes, and impurity ion magnetization angles entering through the projection of total Hamiltonian onto the $S=\frac{5}{2}$ spinor $\ket{\chi\left ( \Theta,\Phi \right )}$. We then add the radial and angular strain energy $\mathcal{E}_{strain}$ to the ground state eigenvalue to calculate the total ground state energy of the system.  

\begin{figure}[!htb]
\begin{center}
\includegraphics[width=3.5in]{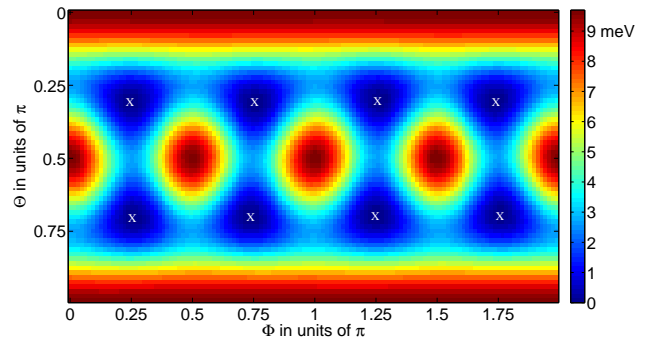}
\end{center}
\caption{Ground state energy of the system as a function of magnetization direction of $\rm{Mn}^{2+}$ ion, parametrized by polar angle $\Theta$ and azimuthal angle $\Phi$, without the inclusion of growth strain. Eight degenerate minima, denoted by ``X'' symbols, corresponding to the magnetization pointing parallel or anti-parallel to the four Mn-As bonds, i.e., along $\langle 111 \rangle$ directions, can be seen. The $\rm{MnAs}_4$ distortion configuration corresponding to the left-most minimum at the top is shown in Fig. \ref{fulldistortion}. Distortion configurations corresponding to other minima have the same structure as in Fig. \ref{fulldistortion}, but with As ions A, B, C and D permuted. The lowest energy barrier separating any two of these minima is about $2.6$ meV.}
\label{Emin_nostrain}
\end{figure}

\begin{figure}[!htb]
\begin{center}
\includegraphics[width=2.8in]{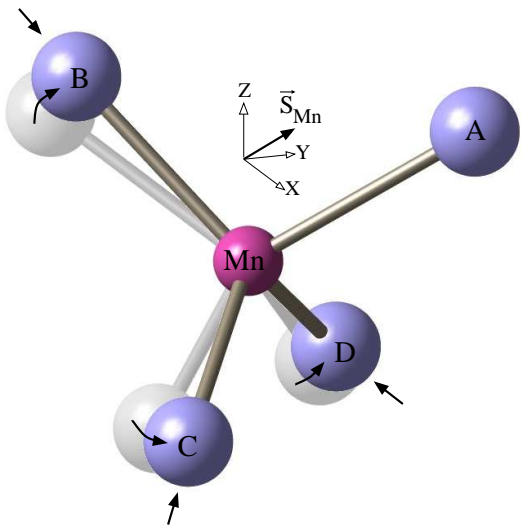}
\end{center}
\caption{Local lattice distortion when the system is ferromagnetic and the magnetization of $\rm{Mn}^{2+}$ ion pointing along Mn-A bond direction, without growth strain effects. Distortions are exaggerated for clarity. The Mn-A bond becomes longer than the other three bonds by about $1.4\%$, and the angles Mn-A bond makes with the other three bonds become smaller by $1.5$ degrees equally. This distortion configuration corresponds to the left-most minimum at the top in Fig. \ref{Emin_nostrain}. The distortion in this ferromagnetic case is opposite to that of the paramagnetic case shown in Fig. \ref{shortdistortion}. In the alternate simplified strain model with fixed As ions, the $\rm{Mn}^{2+}$ moves away from the As ion A.}
\label{fulldistortion}
\end{figure}

We calculate the structural configuration of the system in the ferromagnetic case by minimizing the total ground state energy with respect to the local lattice distortion variables, for all magnetization angles $(\Theta,\Phi)$. Figure \ref{Emin_nostrain} shows the total energy of the system as a function of the magnetization angles. Eight degenerate minima can be observed in the plot, corresponding to the $\rm{Mn}^{2+}$ ion magnetization pointing along and opposite to the directions of the four As ions. In each of the eight minima, the Mn-As bond parallel or anti-parallel to the $\rm{Mn}^{2+}$ ion magnetization direction shrinks less ($3.3\%$ less than $d_0$) than the other three bonds ($4.7\%$ less than $d_0$), and the angles between the longer bond and the other three bonds is smaller than their equilibrium values, by $1.5$ degrees.  The distortions  in the spin-polarized $\rm{MnAs}_4$ configuration move in the opposite direction to that of the paramagnetic case, and is shown in Fig. \ref{fulldistortion}. The minima corresponding to opposite $\rm{Mn}^{2+}$ ion magnetization directions are degenerate and are separated by a finite energy barrier.  Thus, in the ferromagnetic case with no growth strain, magnetic easy axes are \textit{along} $\langle \emph{111} \rangle$ \textit{directions}, when local lattice distortions are included. The reason for the appearance of minima when $\rm{Mn}^{2+}$ ion magnetization points along $\langle 111 \rangle$ directions is that, along these directions, distortions are more effective in maximizing the values of hopping amplitudes between $p$ and $d$ orbitals, thereby, decreasing energy. An EXAFS study\cite{demchenko} of Ga(Mn)As has found that the local structure around the $\rm{Mn}^{2+}$ ion is significantly altered, and also find that, the best fit for their EXAFS oscillations is a model in which the $\rm{Mn}^{2+}$ ion is shifted by $0.1 \rm{\AA}$ from the position of the Ga ion it has replaced. This experiment corroborates our theoretical results. 

\begin{figure}[!htb]
\begin{center}
\includegraphics[width=2.8in]{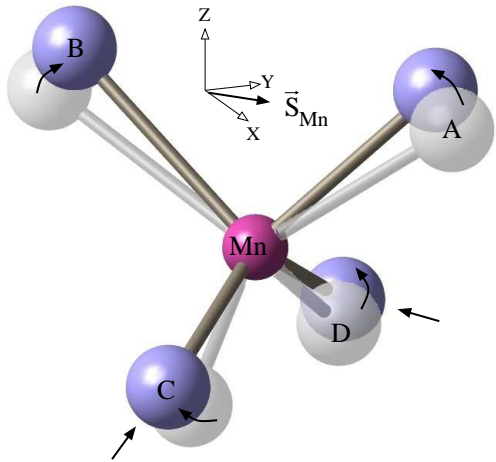}
\end{center}
\caption{Local lattice distortion when magnetization of the $\rm{Mn}^{2+}$ ion points along $\left ( 110 \right )$ direction. All bonds decrease in length, with bonds connecting C and D ions decreasing more ($4.1\%$ less than $d_0$). The angles between C and D increases, whereas, the angle between A and B decreases, both by $1.6$ degrees. Distortions are exaggerated for clarity. The $\rm{Mn}^{2+}$ ion moves towards the x-y plane in the alternate strain model with fixed As ions.}
\label{distortionxy}
\end{figure}

To include compressive growth strain, we remove the heavy-hole basis states, which are higher in energy than light-hole states, from the Hilbert space. The Hilbert space now has just two dimensions, corresponding to the light-hole basis states. The ground state energy, computed by diagonalizing the $2 \times 2$  Hamiltonian matrix, with the $\rm{Mn}^{2+}$ ion magnetization pointing along directions specified by the angles $\Theta$ and $\Phi$, is minimum when the magnetization is in the x-y plane (or $\Theta=\pi/2$), i.e. perpendicular to the growth direction, in accordance with the experiments.
We have also verified the case of tensile strain by including just the heavy-hole doublet, and observed uniaxial, out-of-plane magnetization of $\rm{Mn}^{2+}$ impurity ion. It has to be noted that the demonstrated correspondence between the type of growth strain and magnetization direction happens even in the \textit{absence} of local lattice distortions. This correspondence has to do, as has been explained also by Sawicki \textit{et al}\cite{sawicki1}, with the maximization of hopping possibilities between $\rm{Mn}^{2+}$ ion and the hole orbitals leading to minimization of total energy. In the case of tensile strain, the energetically low-lying heavy-hole states $\ket{J=\frac{3}{2},J_z=\pm\frac{3}{2}}$ have more hopping possibilities with $\ket{S=\frac{5}{2}, S_z=\mp\frac{5}{2}}$ than with other $S_z$ states, thereby lowering the energy of out-of-plane polarized spin states of the impurity ion. Similarly, in the case of compressive strain, the light-hole states $\ket{J=\frac{3}{2},J_z=\pm\frac{1}{2}}$ have more hopping possibilities with $\ket{S=\frac{5}{2}, S_z=\mp\frac{1}{2}}$ than with other $S_z$ states, and hence lower the energy of in-plane spin-polarized states of the impurity ion. 
\begin{figure}[!htb]
\begin{center}
\includegraphics[width=3.5in]{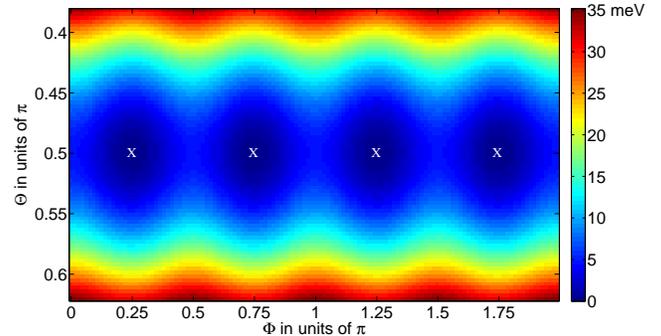}
\end{center}
\caption{Ground state energy of the $\rm{MnAs}_4$ subsystem as a function of the direction of magnetization of the $\rm{Mn}^{2+}$ ion, parametrized by $\Theta$ and $\Phi$, when the system is ferromagnetic and has growth strain that splits heavy and light holes. With the inclusion of only the light holes, minimum ground state energy occurs along $\langle 110 \rangle$ directions ($\Theta=\pi/2$, $\Phi=\pi/4,3\pi/4,5\pi/4,7\pi/4$), as can be seen in the figure, and are denoted by ``X'' symbols. Local lattice distortion configuration corresponding to the first minimum from left is shown in Fig. \ref{distortionxy}. Distortion configurations corresponding to other minima have the same structure as in Fig. \ref{distortionxy}, but with As ions A, B, C and D permuted.} 
\label{Emin_strain}
\end{figure}
\begin{figure}[!htb]
\begin{center}
\includegraphics[width=3.5in]{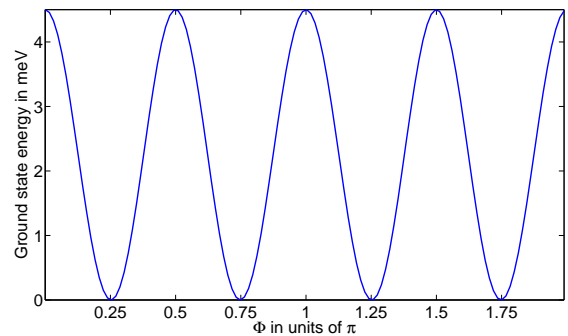}
\end{center}
\caption{Energy as a function of magnetization angle $\Phi$ of the $\rm{Mn}^{2+}$ ion in the x-y plane, in a sample with compressive growth strain. The minima correspond to $\langle 110 \rangle$ magnetization directions ($\Theta=\pi/2$, $\Phi=\pi/4,3\pi/4,5\pi/4,7\pi/4$). The system shows biaxial anisotropy, with the introduction of strain-induced heavy hole - light hole splitting. The energy barrier between two minima is $4.5$ meV.}
\label{Emin_pi_by2}
\end{figure}
Restricting the Hilbert space to just the light-hole doublet results in magnetization pointing in the x-y plane, all directions in the plane being equally possible. Local lattice distortions further reduce the in-plane rotational symmetry and lead to bi-axial magnetic anisotropy. Figure \ref{Emin_strain} shows the total energy of the system as a function of magnetization angles $\Theta$ and $\Phi$. As can be seen in the figure, there are four minima corresponding to the magnetization of $\rm{Mn}^{2+}$ ion pointing along $\langle 110 \rangle$ directions. Variation of energy when the magnetization is rotated in the x-y plane is shown in Fig. \ref{Emin_pi_by2}. Local lattice distortion at one of these minima is shown in Fig. \ref{distortionxy}. 
The figure shows local lattice distortions when magnetization of the $\rm{Mn}^{2+}$ ion points along $\left ( 110 \right )$. In this case, the bonds connecting As ions C and D are shorter ($4.1\%$ less than $d_0$) than that of A and B ($3.5\%$ less than $d_0$), and the angle between C and D increases above $\rm{cos}^{-1}(\frac{1}{3})$, whereas, the angle between A and B decreases, both by about $1.6$ degrees. In the case of $\left ( 1\bar{1}0 \right )$ magnetization, the roles of A and B are replaced by C and D, and vice versa. The $\rm{Mn}^{2+}$ ion moves towards the x-y plane in the equivalent simplified strain model where all other ions are fixed at their equilibrium positions.

\begin{figure}[!htb]
\begin{center}
\includegraphics[width=3.5in]{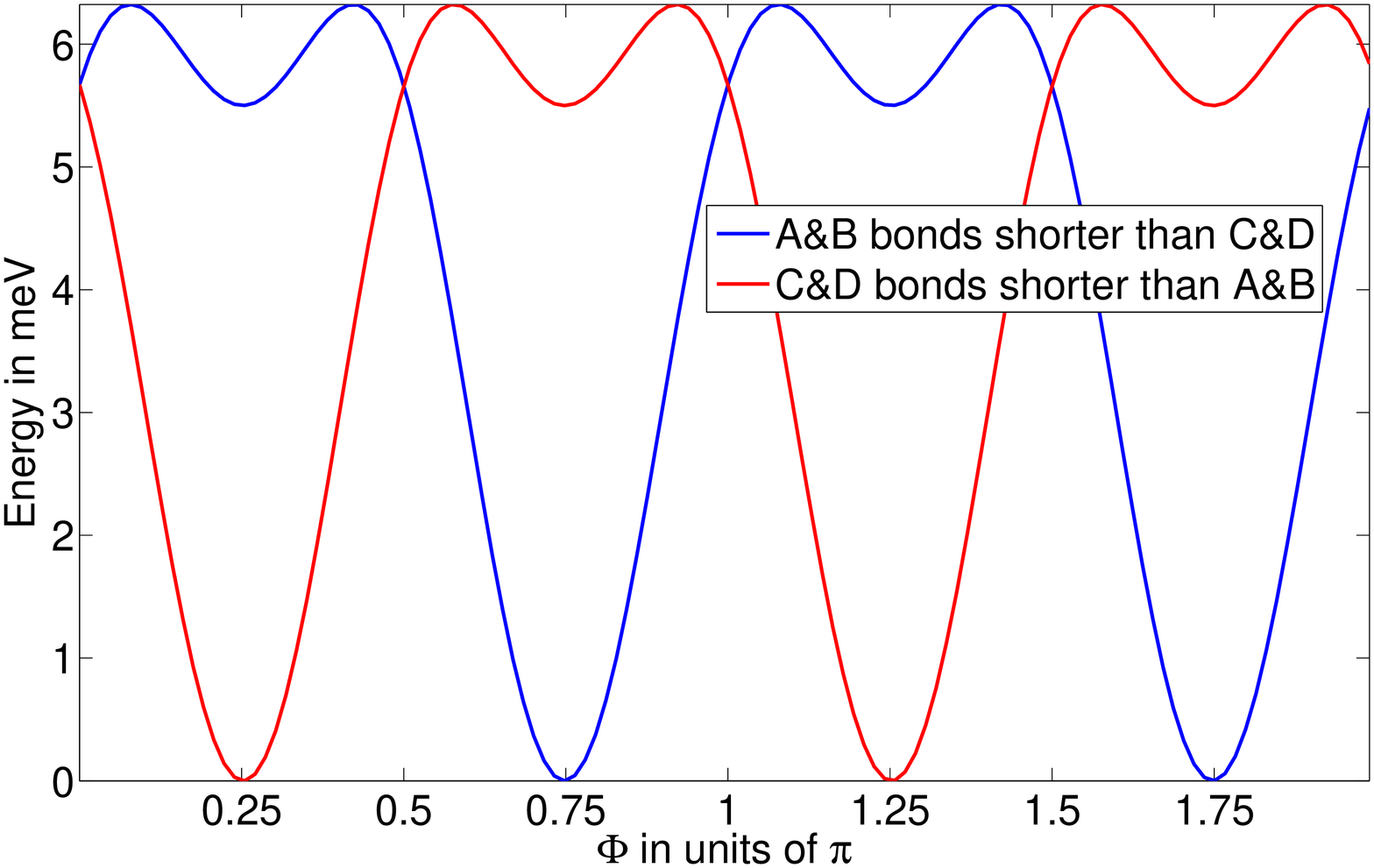}
\end{center}
\caption{Ground state energy as a function of magnetization angle of the $\rm{Mn}^{2+}$ ion in the x-y plane, when compressive growth strain is included, both at the level of band structure by choosing light-holes, and at the level of local lattice structure around the impurity by choosing different equilibrium bond lengths for As ions in the +$z$ and $-z$ directions. The blue curve shows energy variation when the equilibrium lengths of bonds A and B are chosen to be shorter than that of C and D. The system shows uniaxial magnetic anisotropy along $(1\bar{1}0)$ direction. The red curve shows the same scenario with C and D shorter than A and B, resulting in anisotropy along $(110)$ direction.}
\label{Emin_growthstrain}
\end{figure}

We reproduce the experimentally observed uniaxial anisotropy if we include the gradual change in the bond lengths of Ga(Mn)As along $\left ( 001 \right )$ growth direction, from the substrate to the surface. To take this gradual change of bond lengths into account, in the case of compressive strain, we set the equilibrium bond lengths of As ions C and D shorter than that of As ions A and B by $1\%$, as in Fig. \ref{distortionxy}, and minimize the energy with respect to distortion variables. Figure \ref{Emin_growthstrain} shows the variation in total ground state energy as the magnetization of $\rm{Mn}^{2+}$ ion is rotated in the x-y plane. We observe the breaking of the remaining bi-axial symmetry, with the system favoring $\left ( 110 \right )$ direction of magnetization over $\left ( 1\bar{1}0 \right )$ direction. Zemen \textit{et al}\cite{zemen} refer to an experiment where Ga(Mn)As grown on top and bottom surfaces of the same GaAs substrate show perpendicular in-plane magnetizations. This observation can be simply explained within this model once we realize that, for Ga(Mn)As grown on the top surface of GaAs, As ion bonds C and D are shorter, whereas, for Ga(Mn)As grown on the bottom surface, bonds A and B are shorter.  

\section{Summary and conclusions}

In this paper, we have built a many-body tight-binding model for Ga(Mn)As DMS system to explain the experimentally observed uniaxial anisotropy. The model included spin-orbit interaction, intra-atomic Coulomb interaction inside the $\rm{Mn^{2+}}$ impurity ion, hopping between the impurity ion and hole states and local lattice distortions around the impurity ion. The theory applies in the low doping limit (near ${\bf k}=0$, $\Gamma$-point) where we assumed symmetric wavefunction of As $p$ orbitals. We have computationally derived the Heisenberg exchange interaction between the $\rm{Mn^{2+}}$ impurity ion and the hole spins in the distortionless case. Spontaneous tetrahedral symmetry breaking around the impurity ion is observed, even when the system is paramagnetic, when strain energy due to local lattice distortions are included in the Hamiltonian. Magnetization of impurity ion couples to the local lattice distortions through spin-orbit coupling to produce the experimentally observed anisotropies. Under compressive (tensile) strain, we observe in-plane (out-of-plane) magnetic anisotropy, by confining ourselves to light (heavy) holes. Growth strain, or, in other words, the breaking of top-down symmetry, also leads to the breaking of in-plane biaxial symmetry and results in uniaxial anisotropy. 

There is considerable scope for this model to be expanded further, through inclusion of non-zero ${\bf k}$ states of holes, non-zero temperatures and other ionization states of $\rm{Mn^{2+}}$ ion\cite{sapega}, to capture the full complexity of the very interesting anisotropy behavior of Ga(Mn)As, as a function of temperature, hole density and Mn concentration, apart from the strain from substrate. 

\subsection*{Acknowledgements}
We thank Igor Zutic, Karel Vyborny and Rafal Oszwaldowski for helpful suggestions.  We thank Karel Vyborny for his critical reading of the manuscript. 
We gratefully acknowledge the funding from NSF DMR-0907150. 

\section{Appendix}

\subsection{Basis and eigenstates of the acceptor Hamiltonian $\mathcal{H}_a$}\label{holebasis}

In the absence of hopping, spin-orbit coupling and ferromagnetism, the six degenerate many-body basis states of $\mathcal{H}_a$ are denoted as 

\begin{equation}\label{basis}
\begin{aligned}
& \kettt{\uparrow\downarrow}{\uparrow\downarrow}{\downarrow},
\kettt{\uparrow\downarrow}{\uparrow\downarrow}{\downarrow},
\kettt{\uparrow\downarrow}{\downarrow}{\uparrow\downarrow}, \\
& \kettt{\uparrow\downarrow}{\uparrow}{\uparrow\downarrow},
\kettt{\downarrow}{\uparrow\downarrow}{\uparrow\downarrow},
\kettt{\uparrow}{\uparrow\downarrow}{\uparrow\downarrow}, \\
\end{aligned}
\end{equation}
with the three $P$ orbitals being just the $Y_1^{-1}, Y_1^{0}, Y_1^{+1}$ spherical harmonic functions. 
We work out the eigenstates of the hole in the presence of spin-orbit coupling, by diagonalizing $\mathcal{H}_a$ in the basis of Eq. $\eqref{basis}$.

\begin{equation}\label{J3b2}
\begin{aligned}
&\kett{J=\frac{3}{2}}{J_z=\frac{-3}{2}} = \kettt{\uparrow\downarrow}{\uparrow\downarrow}{\downarrow},\\
&\kett{J=\frac{3}{2}}{J_z=\frac{-1}{2}} = \frac{1}{\sqrt{3}} \left( \kettt{\uparrow\downarrow}{\uparrow\downarrow}{\uparrow} - \sqrt{2} \kettt{\uparrow\downarrow}{\downarrow}{\uparrow\downarrow} \right ),\\
&\kett{J=\frac{3}{2}}{J_z=\frac{1}{2}} = \frac{1}{\sqrt{3}} \left( \kettt{\downarrow}{\uparrow\downarrow}{\uparrow\downarrow} - \sqrt{2} \kettt{\uparrow\downarrow}{\uparrow}{\uparrow\downarrow} \right ),\\
&\kett{J=\frac{3}{2}}{J_z=\frac{3}{2}} = \kettt{\uparrow}{\uparrow\downarrow}{\uparrow\downarrow},\\
&\kett{J=\frac{1}{2}}{J_z=\frac{-1}{2}} = \frac{1}{\sqrt{3}} \left( \sqrt{2} \kettt{\uparrow\downarrow}{\uparrow\downarrow}{\uparrow} + \kettt{\uparrow\downarrow}{\downarrow}{\uparrow\downarrow} \right ),\\
&\kett{J=\frac{1}{2}}{J_z=\frac{1}{2}} =  \frac{1}{\sqrt{3}} \left( \sqrt{2} \kettt{\downarrow}{\uparrow\downarrow}{\uparrow\downarrow} + \kettt{\uparrow\downarrow}{\uparrow}{\uparrow\downarrow} \right ).\\
\end{aligned}
\end{equation}

\subsection{Eigenstates of the $\rm{Mn}^{2+}$ impurity Hamiltonian $\mathcal{H}_d$}\label{impuritybasis}

The ground eigenstates of $\mathcal{H}_d$, when five electrons occupy the $d$ orbitals, are given below.
The five $d$ orbitals in the following many-body states are $xy$, $yz$, $zx$, $x^2-y^2$ and $z^2$.  
\begin{equation}\label{s5b2}
\begin{aligned}
\ket{S=\frac{5}{2},S_z=\frac{5}{2}} = \ketfive{\uparrow}{\uparrow}{\uparrow}{\uparrow}{\uparrow} \\
\end{aligned}                   
\end{equation}
\begin{equation}\label{s3b2}
\begin{aligned}
\ket{S=\frac{5}{2}, S_z=\frac{3}{2}} & = \frac{1}{\sqrt{5}} \bigl [ \ketfive{\uparrow}{\uparrow}{\uparrow}{\uparrow}{\downarrow} +            \ketfive{\uparrow}{\uparrow}{\uparrow}{\downarrow}{\uparrow} + \\ 
& \ketfive{\uparrow}{\uparrow}{\downarrow}{\uparrow}{\uparrow} +  \ketfive{\uparrow}{\downarrow}{\uparrow}{\uparrow}{\uparrow} +  \ketfive{\downarrow}{\uparrow}{\uparrow}{\uparrow}{\uparrow} \bigr ] \\
\end{aligned}                   
\end{equation}
\begin{equation}\label{s1b2}
\begin{aligned}
\ket{S=\frac{5}{2}, S_z=\frac{1}{2}} & =  \frac{1}{\sqrt{10}} \bigl [  \ketfive{\uparrow}{\uparrow}{\uparrow}{\downarrow}{\downarrow} +
                               \ketfive{\uparrow}{\uparrow}{\downarrow}{\uparrow}{\downarrow} + \\
                             & \ketfive{\uparrow}{\downarrow}{\uparrow}{\uparrow}{\downarrow} +
                               \ketfive{\downarrow}{\uparrow}{\uparrow}{\uparrow}{\downarrow} + 
                               \ketfive{\uparrow}{\uparrow}{\downarrow}{\downarrow}{\uparrow} + \\
                             & \ketfive{\uparrow}{\downarrow}{\uparrow}{\downarrow}{\uparrow} +
                               \ketfive{\downarrow}{\uparrow}{\uparrow}{\downarrow}{\uparrow} + 
                               \ketfive{\uparrow}{\downarrow}{\downarrow}{\uparrow}{\uparrow} + \\
                             & \ketfive{\downarrow}{\uparrow}{\downarrow}{\uparrow}{\uparrow} + 
                               \ketfive{\downarrow}{\downarrow}{\uparrow}{\uparrow}{\uparrow} \bigr ] \\
\end{aligned}                   
\end{equation}
Negative $S_z$ states are obtained by just flipping the spins in the many-body states from up to down and vice versa.

\subsection{A heuristic understanding of the effective spin Hamiltonian Eq. \eqref{spinint}:}

To motivate the total spin rotational invariant form of the effective Hamiltonian Eq. \eqref{spinint}, we write down the three eigenstates of the total angular momentum operator $\mathbf{F}$, $\ket{F=1,F_z=0,\pm 1}$, in terms of $\ket{S_z,J_z}$:
\begin{equation}\label{Fstates}
\begin{aligned}
{\scriptstyle\ket{F=1,F_z=+1}=}&{\scriptstyle-\sqrt{\frac{1}{2}}\ket{\frac{5}{2},\frac{-3}{2}} + \sqrt{\frac{3}{10}}\ket{\frac{3}{2},\frac{-1}{2}} -\sqrt{\frac{3}{20}}\ket{\frac{1}{2},\frac{1}{2}} + \sqrt{\frac{1}{20}}\ket{\frac{-1}{2},\frac{3}{2}}}, \\
{\scriptstyle\ket{F=1,F_z=0}=} &{\scriptstyle\sqrt{\frac{1}{5}}\ket{\frac{3}{2},\frac{-3}{2}} - \sqrt{\frac{3}{10}}\ket{\frac{1}{2},\frac{-1}{2}} + \sqrt{\frac{3}{10}}\ket{\frac{-1}{2},\frac{1}{2}} - \sqrt{\frac{1}{5}}\ket{\frac{-3}{2},\frac{3}{2}}}, \\
{\scriptstyle\ket{F=1,F_z=-1}=}& {\scriptstyle\sqrt{\frac{1}{20}}\ket{\frac{1}{2},\frac{-3}{2}} - \sqrt{\frac{3}{20}}\ket{\frac{-1}{2},\frac{-1}{2}} +\sqrt{\frac{3}{10}}\ket{\frac{-3}{2},\frac{1}{2}} + \sqrt{\frac{1}{2}}\ket{\frac{-5}{2},\frac{3}{2}}}. \\
\end{aligned}
\end{equation}
To prove that these three eigenstates of $\mathbf{F}$ are also the three lowest eigenstates of the full Hamiltonian $\mathcal{H}$ in the undistorted case, we first note that the off-diagonal elements of $\mathcal{H}_t\frac{1}{(E_0-\mathcal{H}_0)}\mathcal{H}_t$ between the three eigenstates \eqref{Fstates} are zero. This follows from the structure of the hopping amplitudes given in table \ref{hopnostrain}, which shows that the double-hopping process conserves the z-component of the total angular momentum, $F_z=S_z+J_z$. Double-hopping process, irrespective of local lattice distortions, always conserve the z-component of the total \textit{spin} angular momentum, i.e., $\Delta S_z+\Delta s_z^{hole}=0$. In the undistorted case, in addition, z-component of the orbital angular momentum of the final hole state is the same, after double-hopping, as that of the initial hole state, because the $P$ orbital involved in both the hop-out and hop-in processes is the same (please see table \ref{hopnostrain}), and so, $\Delta L_z^{hole}=0$. Therefore, $\Delta F_z = \Delta S_z+\Delta s_z^{hole} + \Delta L_z^{hole} = \Delta S_z + \Delta J_z = 0$. This proves that the off-diagonal elements of the second-order perturbation operator between different $F_z$ eigenstates are zero.   

To prove that the three diagonal matrix elements of $\mathcal{H}_t\frac{1}{(E_0-\mathcal{H}_0)}\mathcal{H}_t$ between $F_z=0,\pm1$ are degenerate, we invoke the rotational symmetry of the ground eigenstates of $\mathcal{H}_a$ and $\mathcal{H}_d$, and the fact that the fifteen hopping amplitudes appear in the Hamiltonian matrix elements only as 
\begin{equation}\label{hopsimple}
\begin{aligned}
T_{m,n} \equiv \sum_{\substack{i=xy,yz,zx,\\x^2\!-\!y^2,z^2}} t_{m,i} \left (t_{n,i} \right)^*.
\end{aligned}
\end{equation} 
In the undistorted case, only the three double hopping amplitudes,  $T_{0,0}$, $T_{+1,+1}$ and $T_{-1,-1}$, are non-zero, and are all equal to $t_0^2$. This results in the three diagonal matrix elements to be the same. Thus, the operator $\mathcal{H}_t\frac{1}{(E_0-\mathcal{H}_0)}\mathcal{H}_t$  is a constant times identity matrix in the sub-space of total angular momentum $\rm{F=1}$, in the case of undistorted $\rm{MnAs}_4$ subsystem.


\begin{thebibliography}{99}
\bibitem{mynote2}
{The reason that the value of $J_0$ is different from the accepted value is because of our inclusion of spin-orbit coupling term in the 
Hamiltonian, whereas it is absent in the model of okabayashi\textit{et al}\cite{okabayashi}, from which we derive the values for our parameters.}
\bibitem{mynote}
{The fact that the expectation value of $\rm{Mn^{2+}}$ spin vanishes under time-reversal symmetric condition makes the assumption of
$\rm{Mn^{2+}}$ spin as a classical vector with a finite magnetic moment untenable under the same condition. Thus this assumption cannot be made in non-ferromagnetic cases or when a system is frustrated, with zero net magnetization.}
\end{thebibliography}
\end{document}